\documentclass[prl,twocolumn,showpacs,superscriptaddress,amssymb]{revtex4}
\usepackage[usenames,dvipsnames]{color}
\usepackage{psfrag,graphicx}
\usepackage{dcolumn}
\usepackage{bm}
\usepackage{amsfonts,amssymb,amsmath}
\usepackage{slashed}

\newcommand{\be}{\begin{equation}}
\newcommand{\ee}{\end{equation}}
\newcommand{\bq}{\begin{eqnarray}}
\newcommand{\eq}{\end{eqnarray}}

\newcommand{\ket}[1]{\left |#1 \right\rangle}
\newcommand{\bra}[1]{\left \langle #1 \right |}

\newcommand{\Eq}[1]{Eq.~(\ref{#1})}

\usepackage{tikz}
\usetikzlibrary[shapes.arrows]
\usetikzlibrary{shapes}
\usetikzlibrary{arrows, decorations.markings}
\tikzstyle{vecArrow} = [thick, decoration={markings,mark=at position
   1 with {\arrow[semithick]{open triangle 60}}},
   double distance=1.4pt, shorten >= 5.5pt,
   preaction = {decorate},
   postaction = {draw,line width=1.4pt, white,shorten >= 4.5pt}]
\tikzstyle{innerWhite} = [semithick, white,line width=1.4pt, shorten >= 4.5pt]

\usepgflibrary{arrows}
\usetikzlibrary{arrows}
\usetikzlibrary{positioning}
\usetikzlibrary{calc}
\usetikzlibrary{decorations.pathmorphing}
\usetikzlibrary{backgrounds}

\definecolor{Blue}{RGB}{0,0,255}
\definecolor{RedOrange}{RGB}{255,165,0}
\definecolor{DarkGreen}{RGB}{0,150,0}
\definecolor{Green}{RGB}{0,210,0}

\definecolor{RedOrange1}{RGB}{255,140,0}
\definecolor{RedOrange2}{RGB}{255,69,0}
\definecolor{Blue1}{RGB}{0,0,205}
\definecolor{Blue2}{RGB}{0,191,255}
\definecolor{fuchsia}{RGB}{152,59,192}

\definecolor{YS}{rgb}{153,153,0}

\begin{document}
\title{Ground State Electroluminescence}
\author{Mauro Cirio}
\affiliation{Interdisciplinary Theoretical Science Research Group (iTHES), RIKEN, Wako-shi, Saitama 351-0198, Japan}
\author{Simone De Liberato}
\affiliation{School of Physics and Astronomy, University of Southampton, Southampton SO17 1BJ, United Kingdom}
\author{Neill Lambert}
\affiliation{CEMS, RIKEN, Saitama, 351-0198, Japan}
\author{Franco Nori}
\affiliation{CEMS, RIKEN, Saitama, 351-0198, Japan}
\affiliation{Department of Physics, University of Michigan, Ann Arbor, MI 48109-1040, USA}
\date{\today}
\begin{abstract}

Electroluminescence, the emission of light in the presence of an electric current, provides information on the allowed electronic transitions of a given system. 
It is commonly used to investigate the physics of  strongly-coupled light-matter systems, whose eigenfrequencies are split by the strong coupling with the photonic field of a cavity.
Here we show that, together with the usual electroluminescence, 
systems in the ultrastrong light-matter coupling regime emit a uniquely quantum radiation when a  flow of current is driven through them.
While standard electroluminescence relies on the population of excited states followed by spontaneous emission, the process we describe herein extracts bound photons by the dressed ground state and it has peculiar features that unequivocally distinguish it from usual electroluminescence.
\end{abstract}
\maketitle

The proper dimensionless parameter to study perturbatively the resonant interaction between light and matter in cavity quantum electrodynamics (QED) is the ratio between vacuum Rabi frequency, $\Omega_R$, and the bare frequency of the excitation, $\omega_C$. When such a quantity, usually referred to as normalised coupling, $\eta\equiv\frac{\Omega_R}{\omega_C}$, becomes non-negligible, higher-order perturbative phenomena start to become observable. This is the so-called ultrastrong-coupling regime, that has been experimentally achieved in a number of solid-state systems \cite{Anappara09,Baust15,Niemczyk10,Muravev11,Geiser12b,Schwartz11,Porer12,Askenazi14,Gubbin14,Gambino14,Maissen14}. The non-perturbative nature of the light-matter coupling in this regime leads to a range of very rich (and as yet not fully understood) phenomenology, including quantum phase transitions \cite{Lambert04,Nataf10b,Baksic13}, modification of energy transport \cite{Orgiu15,Feist15} and light emission  properties \cite{DeLiberato14,Ripoll15,Bamba15}, or the appearance of cavity-assisted chemical and thermodynamic effects \cite{Hutchison12,Hutchison13,Galego15,Cwik15}.
One of the main consequences of the ultrastrong-coupling regime is to modify the ground state of the system, such that it becomes a multimode squeezed state, containing a finite population of bound excitations that can only be observed if the system parameters are nonadiabatically modulated in time \cite{Ciuti05,Dodonov06,DeLiberato07,DeLiberato09,Agnesi09,Faccio11,Carusotto12,Auer12}.
This quantum vacuum radiation, that has strong similarities with the dynamical Casimir
effect \cite{Lambrecht07,Johansson09,Johansson10,Wilson11,Nation12,Johansson13}, is a hallmark of the quantum nature of
the light-matter interaction, and it can provide insights into the quantum nature of the ground state \cite{Lolli15}.\\
Here we investigate a unique quantum electrodynamical effect wherein one can observe the photonic excitations bound in the ground state when an electrical current is driven through the system. We call this effect {\em ground state electroluminescence} (GSE), in order to distinguish it from the standard polaritonic electroluminescence observable when a current passes in a strongly-coupled light-matter system \cite{Khalifa08,Tsintzos08,Sapienza08,Lodden11,Jouy12,Geiser12,Gubbin14,Astafiev07,Liu14,Stockklauser15,DeLiberato08,DeLiberato09b}.
We will see that such a GSE is not only intense enough to be observable in some near-future solid-state cavity QED experiments, but it has also unique features that clearly distinguish it from the usual polaritonic electroluminescence.


\begin{figure}[t!]
\centering
\begin{tikzpicture}
\node [font=\large,scale=.8,style=black] at  (-4.85,3.2) {(a)};
\node [font=\large,scale=.8,style=black] at  (-0.45,3.2) {(b)};
\node [font=\large,scale=.8,style=black] at  (-4.85,-1) {(c)};
\node [font=\large,scale=.8,style=black] at  (-0.45,-1) {(d)};

\node at (-2.8,1.3){\scalebox{0.5}{\begin{tikzpicture}[>=latex,thick,font=\normalsize]
\node at (0,2.5)[rectangle,draw=black!0,fill=black!0,thick,inner sep=0pt,minimum width=2.8cm,minimum height=.01cm] (invisible) {};
\node [font=\large,scale=1.3,style=black,opacity=0] at  ($($(invisible)$)-(0.8,-0.35)$) {$\ket{+}$};
\node at (0,2)[rectangle,draw=black!100,fill=black!100,thick,inner sep=0pt,minimum width=2.8cm,minimum height=.01cm] (a) {};
\node at (0,-1.2)[rectangle,draw=black!100,fill=black!100,thick,inner sep=0pt,minimum width=2.8cm,minimum height=.01cm] (b) {};
\node at (b.east)[anchor=west,rectangle,draw=black!0,fill=black!100,thick,inner sep=0pt,minimum width=3.cm,minimum height=.01cm, dashed] (bbb) {};
\node at (0,-4.5)[rectangle,draw=black!100,fill=black!100,thick,inner sep=0pt,minimum width=2.8cm,minimum height=.01cm] (c) {};
\node at (c.east)[anchor=west,rectangle,draw=black!0,fill=black!100,thick,inner sep=0pt,minimum width=3.cm,minimum height=.01cm, dashed] (cc) {};
\node at ($($(c.east)$)+(0,1.6)$)[anchor=west,rectangle,draw=black!100,fill=black!100,thick,inner sep=0pt,minimum width=2.8cm,minimum height=.01cm] (d) {};

\node at (d.west)[anchor=east,rectangle,draw=black!0,fill=black!100,thick,inner sep=0pt,minimum width=2.8cm,minimum height=.01cm,dashed] (dd){};

\node at ($($(b.east)$)+(0,1.6)$)[anchor=west,rectangle,draw=black!100,fill=black!100,thick,inner sep=0pt,minimum width=2.8cm,minimum height=.01cm] (f) {};

\node at (f.west)[anchor=east,rectangle,draw=black!0,fill=black!100,thick,inner sep=0pt,minimum width=2.8cm,minimum height=.01cm,dashed] (ff) {};

\draw [-,black,shorten >=2pt,decoration={markings, mark=at position 0.999 with {\arrow[scale=1.5]{>}}; }, postaction={decorate},decoration={markings, mark=at position 0.001 with {\arrow[scale=1.5]{<}}; }, postaction={decorate}] ($($(c)$)-(-0.6,-0.28)$) -- ($($(dd)$)-(-0.6,0.)$); 
\node [font=\large,scale=1.3,style=black] at  ($0.5*($($($(c)$)-(-0.6,0)$) + ($($(dd)$)-(-0.6,0)$)$)+(0.4,0)$) {$\omega_C$};
\draw [-,black,shorten >=2pt,decoration={markings, mark=at position 0.999 with {\arrow[scale=1.5]{>}}; }, postaction={decorate},decoration={markings, mark=at position 0.001 with {\arrow[scale=1.5]{<}}; }, postaction={decorate}] ($($(c)$)-(0,-0.28)$) -- ($($(b)$)-(0,0)$); 
\node [font=\large,scale=1.3,style=black] at  ($0.5*($($($(c)$)-(-0,0)$) + ($($(b)$)-(-0.,0)$)$)+(0.35,0.4)$) {$\omega_s$};

\draw [-,black,shorten >=2pt,decoration={markings, mark=at position 0.999 with {\arrow[scale=1.5]{>}}; }, postaction={decorate},decoration={markings, mark=at position 0.001 with {\arrow[scale=1.5]{<}}; }, postaction={decorate}] ($($(b)$)-(-0.6,-0.28)$) -- ($($(ff)$)-(-0.6,0)$); 
\node [font=\large,scale=1.3,style=black] at  ($0.5*($($($(b)$)-(-0.6,0)$) + ($($(ff)$)-(-0.6,0)$)$)+(0.4,0)$) {$\omega_C$};
\draw [-,black,shorten >=2pt,decoration={markings, mark=at position 0.999 with {\arrow[scale=1.5]{>}}; }, postaction={decorate},decoration={markings, mark=at position 0.001 with {\arrow[scale=1.5]{<}}; }, postaction={decorate}] ($($(b)$)-(0,-0.28)$) -- ($($(a)$)-(0,0)$); 
\node [font=\large,scale=1.3,style=black] at  ($0.5*($($($(b)$)-(-0,0)$) + ($($(a)$)-(-0.,0)$)$)+(0.35,0.4)$) {$\omega_e$};

\draw [->,red,line width=1.6pt] ($($(c)$)-(2.15,0)$) -- ($($(b)$)-(2.15,0)$); 
\draw [<-,cyan,line width=1.6pt] ($($(c)$)-(2,0)$) -- ($($(b)$)-(2,0)$); 
\node [font=\large,scale=1.3,style=black] at   ($0.5*($($($(c)$)-(2,0)$) + ($($(b)$)-(2.,0)$)$)+(0.8,1.)$){$\Gamma_{\text{in/out}}$};

\draw [->,red,line width=1.6pt] ($($(c)$)-(2.6,0)$) -- ($($(a)$)-(2.6,0)$); 
\draw [<-,cyan,line width=1.6pt] ($($(c)$)-(2.45,0)$) -- ($($(a)$)-(2.45,0)$); 
\node [font=\large,scale=1.3,style=black] at   ($0.5*($($($(c)$)-(2,0)$) + ($($(a)$)-(2.,0)$)$)+(0.4,2.7)$){$\Gamma_{\text{in/out}}$};

\draw [->,red,line width=1.6pt] ($($(d)$)+(1.5,0)$) -- ($($(f)$)+(1.5,0)$); 
\draw [<-,cyan,,line width=1.6pt] ($($(d)$)+(1.65,0)$) -- ($($(f)$)+(1.65,0)$); 
\node [font=\large,scale=1.3,style=black] at   ($0.5*($($($(d)$)+(2,0)$) + ($($(f)$)+(2.,0)$)$)+(-1.3,-0.5)$){$\Gamma_{\text{in/out}}$};

\node [scale=1, single  arrow, draw,rotate=-135,draw=DarkGreen!100,fill=Green!40] at ($($(d)$)-(1.,0.7)$) (gammaCav) {~~~~~~~~}; 
\node [font=\large,scale=1.3,style=black] at   ($($(gammaCav)$)+(1.,0.)$){$\Gamma_{\text{cav}}$};

\node [scale=1, single  arrow, draw,rotate=-135,draw=DarkGreen!100,fill=Green!40] at ($($(f)$)-(1.,0.7)$) (gammaCav2) {~~~~~~~~}; 
\node [font=\large,scale=1.3,style=black] at   ($($(gammaCav2)$)+(1.,0.)$){$\Gamma_{\text{cav}}$};

\node [font=\large,scale=1.3,style=black] at  ($($(c)$)-(0.8,-0.35)$) {$\ket{s,0}$};
\node [font=\large,scale=1.3,style=black] at  ($($(b)$)-(0.8,-0.35)$) {$\ket{g,0}$};
\node [font=\large,scale=1.3,style=black] at  ($($(a)$)-(0.8,-0.35)$) {$\ket{e,0}$};
\node [font=\large,scale=1.3,style=black] at  ($($(d)$)-(0.8,-0.35)$) {$\ket{s,1}$};
\node [font=\large,scale=1.3,style=black] at  ($($(f)$)-(0.8,-0.35)$) {$\ket{g,1}$};

\end{tikzpicture}}};


\node at (1.15,1.3){\scalebox{0.5}{\begin{tikzpicture}[>=latex,thick,font=\normalsize]
\node at (0,2.5)[rectangle,draw=black!100,fill=black!100,thick,inner sep=0pt,minimum width=2.8cm,minimum height=.01cm] (aa) {};

\node at (aa.east)[anchor=west,rectangle,draw=black!0,fill=black!100,thick,inner sep=0pt,minimum width=2.8cm,minimum height=.01cm,dashed] (aa2) {};
\node at (0,2.)[rectangle,draw=black!00,fill=black!00,thick,inner sep=0pt,minimum width=2.8cm,minimum height=.01cm] (a) {};
\node at (0,-2.)[rectangle,draw=black!100,fill=black!100,thick,inner sep=0pt,minimum width=2.8cm,minimum height=.01cm] (b) {};

\node at (b.east)[anchor=west,rectangle,draw=black!0,fill=black!100,thick,inner sep=0pt,minimum width=2.8cm,minimum height=.01cm,dashed] (b2) {};

\node at (0,-1.2)[rectangle,thick,inner sep=0pt,minimum width=2.8cm,minimum height=.01cm,draw=black!0,fill=black!100,dashed] (bOld) {};

\node at (bOld.west)[anchor=east,rectangle,draw=black!0,fill=black!100,inner sep=0pt,minimum width=2.9cm,minimum height=.01cm,dashed] (bbOld) {};

\node at (0,-4.5)[rectangle,draw=black!100,fill=black!100,thick,inner sep=0pt,minimum width=2.8cm,minimum height=.01cm] (c) {};
\node at (c.west)[anchor=east,rectangle,draw=black!0,fill=black!100,inner sep=0pt,minimum width=3cm,minimum height=.01cm,dashed] (cc) {};

\node at ($($(c.east)$)+(0,1.6)$)[anchor=west,rectangle,draw=black!100,fill=black!100,thick,inner sep=0pt,minimum width=2.8cm,minimum height=.01cm] (d) {};

\node at (d.west)[anchor=east,rectangle,draw=black!0,fill=black!100,thick,inner sep=0pt,minimum width=2.8cm,minimum height=.01cm,dashed] (dd){};

\node at ($($(b.east)$)+(0,3.6)$)[anchor=west,rectangle,draw=black!00,fill=black!00,thick,inner sep=0pt,minimum width=2.8cm,minimum height=.01cm] (f) {};

\node at (f.west)[anchor=east,rectangle,draw=black!100,fill=black!100,thick,inner sep=0pt,minimum width=2.8cm,minimum height=.01cm] (ff) {};

\draw [-,black,shorten >=2pt,decoration={markings, mark=at position 0.999 with {\arrow[scale=1.5]{>}}; }, postaction={decorate},decoration={markings, mark=at position 0.001 with {\arrow[scale=1.5]{<}}; }, postaction={decorate}] ($($(c)$)-(-0.6,-0.28)$) -- ($($(dd)$)-(-0.6,0)$); 
\node [font=\large,scale=1.3,style=black] at  ($0.5*($($($(c)$)-(-0.6,0)$) + ($($(dd)$)-(-0.6,0)$)$)+(0.4,0)$) {$\omega_C$};
\draw [-,black,shorten >=2pt,decoration={markings, mark=at position 0.999 with {\arrow[scale=1.5]{>}}; }, postaction={decorate},decoration={markings, mark=at position 0.001 with {\arrow[scale=1.5]{<}}; }, postaction={decorate}] ($($(bOld)$)-(1.3,0.28)$) -- ($($(b)$)-(1.3,-0.)$); 
\node [font=\large,scale=1.3,style=black] at  ($0.5*($($($(bOld)$)-(1.3,0)$) + ($($(b)$)-(1.3,0)$)$)-(0.5,0)$) {$\omega_G$};

\draw [-,black,shorten >=2pt,decoration={markings, mark=at position 0.999 with {\arrow[scale=1.5]{>}}; }, postaction={decorate},decoration={markings, mark=at position 0.001 with {\arrow[scale=1.5]{<}}; }, postaction={decorate}] ($($(b)$)-(0.1,-0.28)$) -- ($($(ff)$)-(0.1,0)$); 
\node [font=\large,scale=1.3,style=black] at  ($0.5*($($($(b)$)-(0.1,0)$) + ($($(ff)$)-(0.1,0)$)$)+(0.4,0)$) {$\omega_-$};
\draw [-,black,shorten >=2pt,decoration={markings, mark=at position 0.999 with {\arrow[scale=1.5]{>}}; }, postaction={decorate},decoration={markings, mark=at position 0.001 with {\arrow[scale=1.5]{<}}; }, postaction={decorate}] ($($(b)$)-(-2.25,-0.28)$) -- ($($(aa)$)-(-2.25,0)$); 
\node [font=\large,scale=1.3,style=black] at  ($0.5*($($($(b)$)-(-2.2,0)$) + ($($(aa)$)-(-2.2,0)$)$)+(0.45,0)$) {$\omega_+$};

\draw [-,black,shorten >=2pt,decoration={markings, mark=at position 0.999 with {\arrow[scale=1.5]{>}}; }, postaction={decorate},decoration={markings, mark=at position 0.001 with {\arrow[scale=1.5]{<}}; }, postaction={decorate}] ($($(ff)$)-(0.1,-0.28)$) -- ($($(aa)$)-(0.1,0)$); 
\node [font=\large,scale=1.3,style=black] at  ($0.5*($($($(ff)$)-(0.1,0)$) + ($($(aa)$)-(0.1,0)$)$)+(0.8,0)$) {$\;\simeq2\Omega_R$};

\draw [<-,cyan,line width=1.6pt] ($($(d)$)+(0,0.05)$) -- ($($(b2)$)-(0,0.05)$); 
\node [font=\large,scale=1.3,style=black] at  ($0.5*($($($(d)$)-(0,0)$) + ($($(b2)$)-(0,0.05)$)$)+(0.75,0.05)$) {$\Gamma^{G\rightarrow 1}_{\text{out}}$};

\draw [->,red,line width=1.6pt] ($($(c)$)-(2.15,0)$) -- ($($(b)$)-(2.15,0)$); 
\draw [<-,cyan,line width=1.6pt] ($($(c)$)-(2,0)$) -- ($($(b)$)-(2,0)$); 
\node [font=\large,scale=1.3,style=black] at   ($0.5*($($($(c)$)-(2,0)$) + ($($(b)$)-(2.,0)$)$)+(0.7,-0.1)$){$\Gamma^{0\rightarrow G}_{\text{in}}$};
\node [font=\large,scale=1.3,style=black] at   ($0.5*($($($(c)$)-(2,0)$) + ($($(b)$)-(2.,0)$)$)+(0.7,0.8)$){$\Gamma^{G\rightarrow 0}_{\text{out}}$};

\draw [->,red,line width=1.6pt] ($($(c)$)-(2.55,0)$) -- ($($(aa)$)-(2.55,0)$); 
\node [font=\large,scale=1.3,style=black] at   ($0.5*($($($(c)$)-(2,0)$) + ($($(aa)$)-(2.,0)$)$)+(-0.05,3.7)$){$\Gamma^+_{\text{in}}$};

\draw [->,red,line width=1.6pt] ($($(c)$)-(2.4,0)$) -- ($($(ff)$)-(2.4,0)$); 
\node [font=\large,scale=1.3,style=black] at   ($0.5*($($($(c)$)-(2,3)$) + ($($(ff)$)-(2.,3)$)$)+(0.1,6.2)$){$\Gamma^-_{\text{in}}$};

\node [scale=1, single  arrow, draw,rotate=-135,draw=DarkGreen!100,fill=Green!40] at ($($(d)$)-(1.,0.7)$) (gammaCav) {~~~~~~~~}; 
\node [font=\large,scale=1.3,style=black] at   ($($(gammaCav)$)+(1.,0.)$){$\Gamma_{\text{cav}}$};

\node [scale=1, single  arrow, draw,rotate=-90,draw=DarkGreen!100,fill=Green!40] at ($($(aa2)$)+(.4,-2.25)$) (gammaCav2) {~~~~~~~~~~~~~~~~~~~~~~~~~~~~~~}; 
\node [font=\large,scale=1.3,style=black] at   ($($(gammaCav2)$)+(.7,0.05)$){$\Gamma^+_{\text{cav}}$};

\node [scale=1, single  arrow, draw,rotate=-90,draw=DarkGreen!100,fill=Green!40] at ($($(ff)$)+(1,-1.8)$) (gammaCav3) {~~~~~~~~~~~~~~~~~~~~~~}; 
\node [font=\large,scale=1.3,style=black] at   ($($(gammaCav3)$)+(.7,0.05)$){$\Gamma^-_{\text{cav}}$};

\node [font=\large,scale=1.3,style=black] at  ($($(c)$)-(0.8,-0.35)$) {$\ket{s,0}$};
\node [font=\large,scale=1.3,style=black] at  ($($(b)$)-(0.8,-0.35)$) {$\ket{G}$};
\node [font=\large,scale=1.3,style=black] at  ($($(aa)$)-(0.8,-0.35)$) {$\ket{+}$};
\node [font=\large,scale=1.3,style=black] at  ($($(d)$)-(0.8,-0.35)$) {$\ket{s,1}$};
\node [font=\large,scale=1.3,style=black] at  ($($(ff)$)-(0.8,-0.35)$) {$\ket{-}$};

\end{tikzpicture}}};

\node at (-3.2,-2.5){\scalebox{0.55}{\begin{tikzpicture}[node distance=-0.mm,>=latex,thick,font=\normalsize]
\node at (0,1.)[rectangle,draw=black!100,fill=black!10,thick,inner sep=0pt,minimum width=2.5cm,minimum height=.8cm](cavity1){};
\node at (cavity1.south)[ellipse,draw=black!100,fill=white!100,thick,inner sep=0pt,minimum width=1.9cm,minimum height=1cm]{};
\node at (cavity1.south)[anchor=north,rectangle,draw=white!100,fill=white!100,thick,inner sep=0pt,minimum width=2.5cm,minimum height=1cm]{};

\node at (0,-3.)[rectangle,draw=black!100,fill=black!10,thick,inner sep=0pt,minimum width=2.5cm,minimum height=.8cm](cavity2){};
\node at (cavity2.north)[ellipse,draw=black!100,fill=white!100,thick,inner sep=0pt,minimum width=1.9cm,minimum height=1cm]{};
\node at (cavity2.north)[anchor=south,rectangle,draw=white!100,fill=white!50,thick,inner sep=0pt,minimum width=2.5cm,minimum height=1cm]{};

\node at (0,0)[rectangle,draw=black!100,fill=black!100,thick,inner sep=0pt,minimum width=1.8cm,minimum height=.02cm] (a) {};
\node at (0,-1.7)[rectangle,draw=black!0,fill=black!0,thick,inner sep=0pt,minimum width=1.8cm,minimum height=.02cm] (b) {};
\node at (0,-2)[rectangle,draw=black!100,fill=black!100,thick,inner sep=0pt,minimum width=1.8cm,minimum height=.02cm] {};

\node [anchor=center,font=\large,scale=1.3,style=black]  at ($($(b.center)$)-(0,0.8)$) {$\ket{G}$};
\node [anchor=center,font=\large,scale=1.3,style=black]  at ($($(a.center)$)+(0,0.4)$) {$\ket{\pm}$};

\node at ($($(a)$)-(3.,1.15)$)[rectangle,draw=black!100,fill=black!00,thick,inner sep=0pt,minimum width=0.3cm,minimum height=4cm] (aa) {};
\node at ($(aa.south)$) [anchor=south,rectangle,draw=black!100,fill=black!60,thick,inner sep=0pt,minimum width=0.3cm,minimum height=3.5cm]  {};
\node at ($($(a)$)-(-3.,1.15)$)[rectangle,draw=black!100,fill=black!00,thick,inner sep=0pt,minimum width=0.3cm,minimum height=4cm]  (bb){};
\node at ($(bb.south)$) [anchor=south,rectangle,draw=black!100,fill=black!60,thick,inner sep=0pt,minimum width=0.3cm,minimum height=0.5cm]  {};
\node [scale=1.2, single  arrow, draw,rotate=0,draw=RedOrange2!100,fill=RedOrange!40] at ($($(a)$)-(2.05,0)$) (arrow1) {$\Gamma^{0\rightarrow\pm}_{\text{in}}$}; 
\node [anchor=center,scale=1.2, single  arrow, draw,rotate=-90,draw=RedOrange2!100,fill=RedOrange!40] at ($0.5*($(a)$)+($0.5*(b)$)+(0,0.1)$) {~~\rotatebox{90}{$\Gamma^{\pm}_{\text{cav}}$}~~}; 
\node [scale=1.2, single  arrow, draw,rotate=0,draw=RedOrange2!100,,fill=RedOrange!40] at ($($(b.east)$)+(.8,-0.3)$) (arrow2) {$\Gamma^{G\rightarrow0}_{\text{out}}$};

 \node [anchor=center,font=\large,scale=1.3,style=black]  at (2,0.7) {$\hbar\omega_\pm$}; 

\node [anchor=west] at ($0.5*($(a)$)+($0.5*(b)$)+(0.45,0.65)$){\scalebox{1}{\begin{tikzpicture}[
  >=stealth',
  pos=0,
  photon/.style={decorate,decoration={snake,post length=1.9mm}}
]
 \draw[->,photon,color=YS!70!black,line width=1.6pt] (0,0) --(1,1.2) {} ;
\end{tikzpicture}}};

\end{tikzpicture}}};


\node at (1.2,-2.5){\scalebox{0.55}{\begin{tikzpicture}[node distance=-0.mm,>=latex,thick,font=\normalsize]

\node at (0,1.)[rectangle,draw=black!100,fill=black!10,thick,inner sep=0pt,minimum width=2.5cm,minimum height=.8cm](cavity1){};
\node at (cavity1.south)[ellipse,draw=black!100,fill=white!100,thick,inner sep=0pt,minimum width=1.9cm,minimum height=1cm]{};
\node at (cavity1.south)[anchor=north,rectangle,draw=white!100,fill=white!100,thick,inner sep=0pt,minimum width=2.5cm,minimum height=1cm]{};

\node at (0,-3.)[rectangle,draw=black!100,fill=black!10,thick,inner sep=0pt,minimum width=2.5cm,minimum height=.8cm](cavity2){};
\node at (cavity2.north)[ellipse,draw=black!100,fill=white!100,thick,inner sep=0pt,minimum width=1.9cm,minimum height=1cm]{};
\node at (cavity2.north)[anchor=south,rectangle,draw=white!100,fill=white!50,thick,inner sep=0pt,minimum width=2.5cm,minimum height=1cm]{};

\node at (0,0)[rectangle,draw=black!100,fill=black!100,thick,inner sep=0pt,minimum width=1.8cm,minimum height=.02cm] (a) {};
\node at (0,-1.7)[rectangle,draw=black!0,fill=black!0,thick,inner sep=0pt,minimum width=1.8cm,minimum height=.02cm] (b) {};
\node at (0,-2)[rectangle,draw=black!100,fill=black!100,thick,inner sep=0pt,minimum width=1.8cm,minimum height=.02cm] {};

\node [anchor=center,font=\large,scale=1.3,style=black]  at ($($(b.center)$)-(0,0.8)$) {$\ket{G}$};
\node [anchor=center,font=\large,scale=1.3,style=black]  at ($($(a.center)$)+(0,0.4)$) {$\ket{\pm}$};

\node at ($($(a)$)-(3,1.15)$)[rectangle,draw=black!100,fill=black!00,thick,inner sep=0pt,minimum width=0.3cm,minimum height=4cm] (aa) {};
\node at ($(aa.south)$) [anchor=south,rectangle,draw=black!100,fill=black!60,thick,inner sep=0pt,minimum width=0.3cm,minimum height=1.5cm]  {};
\node at ($($(a)$)-(-3,1.15)$)[rectangle,draw=black!100,fill=black!00,thick,inner sep=0pt,minimum width=0.3cm,minimum height=4cm]  (bb){};
\node at ($(bb.south)$) [anchor=south,rectangle,draw=black!100,fill=black!60,thick,inner sep=0pt,minimum width=0.3cm,minimum height=0.5cm]  {};

\node [scale=1.2, single  arrow, draw,rotate=0,draw=Blue1!100,fill=Blue!10] at ($($(b.east)$)+(0.8,-0.3)$) {$\Gamma^{G\rightarrow1}_{\text{out}}$}; 

\node [scale=1.2, single  arrow, draw,rotate=0,draw=Blue1!100,fill=Blue!10] at ($($(b)$)-(2.05,0.3)$) (arrow10) {$\Gamma^{0\rightarrow G}_{\text{in}}$};

 \node [anchor=center,font=\large,scale=1.3,style=black]  at (2.2,-0) {$\hbar\omega_C$}; 
 
 \node [anchor=west] at ($0.5*($(a)$)+($0.5*(b)$)+(1.3,-0.1)$){\scalebox{1}{\begin{tikzpicture}[
  >=stealth',
  pos=0,
  photon/.style={decorate,decoration={snake,post length=1.9mm}}
]
 \draw[->,photon,color=YS!70!black,line width=1.6pt] (0,0) --(1,1.2) {} ;

\end{tikzpicture}}};

\end{tikzpicture}}};


 \end{tikzpicture}  
 \caption{\label{Fig1} (a) Energy diagram for the uncoupled system ($\Omega_R=0$). Photons can leave the cavity at a rate $\Gamma_{\text{cav}}$ and electrons can populate/leave the system at rates $\Gamma_{\text{in/out}}$. (b) Diagram levels for the strongly coupled regime at resonance $\omega_C=\omega_e$. (c) Schematic showing standard electroluminescent emission. (d) Schematic showing the dominant GSE process.}
\end{figure}
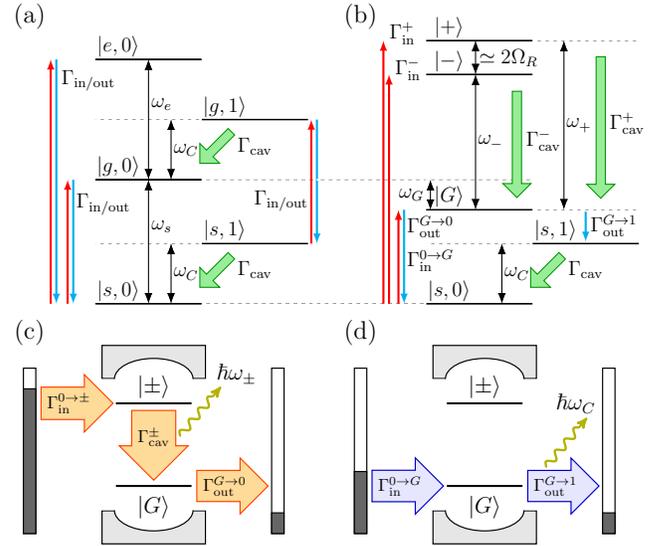

Our aim is to present a novel QED effect, a priori relevant for many of the systems
in which the USC regime has been observed. For this reason we will keep the discussion as generic and system-independent as possible, investigating the simplest model presenting GSE effects. A discussion on possible implementations will be discussed later.
We will thus consider a toy model exhibiting GSE: a two-level electronic system ultrastrongly coupled to a single mode of a  photonic resonator. When a single electron is present, this is just the Rabi model, one of the simplest models used to investigate the physics of strong light-matter coupling. The important element we add here is that, given that we are interested in electroluminescence, we will consider the system to be electronically open: i.e., we will account for the possibility of  electrons to tunnel in and out of the structure. While it is not necessary for our theory, in order to simplify the already rather heavy notation, we will neglect the possibility to have a doubly-occupied state. Depending on the specific implementation, this assumption can be physically justified assuming the system to be in a Coulomb-blockade regime.
Thus, for this electronic system we only consider a 3D Hilbert space spanned by the empty state $\ket{s}$, and the two singly-occupied states $\ket{g}$ and $\ket{e}$, with Hamiltonian
\begin{eqnarray}
\label{H}
H&=&\hbar\omega_C a^\dagger a+
\hbar\omega_e\ket{e}\bra{e}-\hbar\omega_s\ket{s}\bra{s}
\nonumber\\
&&+\Omega_R(a+a^\dagger)(\ket{e}\bra{g}+\ket{g}\bra{e}),\nonumber
\end{eqnarray}
where $a$ is the annihilation operator of the photonic mode of energy $\hbar\omega_C$, and $\hbar\omega_e$ and $\hbar\omega_s$ the energies of the  
$g\rightarrow e$ and $s\rightarrow g$ transitions. For definiteness, hereafter we will limit ourselves to the resonant case $\omega_e=\omega_C$.
As the Hamiltonian in \Eq{H} conserves the number of electrons, it can be diagonalised separately in the zero and one electron sectors. In the first sector the Hamiltonian is diagonal in the $n$-photons states $\ket{s,n}$, of energy $n\hbar\omega_C-\hbar\omega_s$; while in the second sector we recover exactly the Hamiltonian of a quantum Rabi model \cite{Braak11}. We will call respectively $\ket{G}$ and $\ket{\pm}$ its ground state and first polaritonic doublet, with respective energies $\hbar\omega_G$ and $\hbar\omega_G+\hbar\omega_{\pm}$. For the sake of clarity, we will neglect higher-lying states and temperature effects, as both are inessential to GSE.
In order to be able to investigate electroluminescence, we  couple such a system to two external electronic reservoirs. Fixing the chemical potential of one of them to $-\hbar\omega_s$, so that it acts only as a sink for  electrons, the voltage across the system is described by the chemical potential $\mu$ of the other reservoir, leading to the effective electronic injection and extraction rates $\Gamma_{\text{in/out}}$. We also couple the cavity itself to a reservoir, describing the extra-cavity photonic modes in which the luminescence is emitted, leading to a photonic lifetime $\Gamma_{\text{cav}}$. A diagram of the relevant levels and transition rates for vanishing couplings is in Fig.~\ref{Fig1}(a).

Note that we will consider strongly-coupled systems, whose eigenstates are linear superpositions of different electronic and photonic excitations. Therefore, the bare electronic and photonic lifetimes will give rise to different transition rates between different pairs of dressed states, that can be calculated using the different methods that have been developed to deal with strongly-coupled open systems \cite{DeLiberato09,Beaudoin11,Ridolfo12,DeLiberato14b}. 
The  levels and transition rates for strong couplings is in Fig.~\ref{Fig1}(b). 
Limiting ourselves to the low-lying states that we will consider here, we thus define $\Gamma_{\text{in}}^{n\rightarrow j}$ and $\Gamma_{\text{out}}^{j \rightarrow n}$ as the electronic injection and extraction rates to and from the states $\ket{s,n}$ and $\ket{j}$, $j\in\{ G,\pm \}$, and $ \Gamma_{\text{cav}}^{\pm}$ as the photonic transition rates from $\ket{\pm}$ to $\ket{G}$.
Notice that in the zero temperature case we are considering here, electron tunnelling can couple two states only if $\mu$ is larger than their energy mismatch, that is 
\begin{eqnarray}
\label{Inrate}
\Gamma_{\text{in}}^{n\rightarrow G}&\propto&\Theta(\mu+n\hbar\omega_C-\hbar\omega_G)\nonumber\\ \Gamma_{\text{in}}^{n\rightarrow \pm,}&\propto&\Theta(\mu+n\hbar\omega_C-\hbar\omega_G-\hbar\omega_{\pm}),
\end{eqnarray}
where $\Theta$ is the Heaviside function.

In order to introduce our discussion of GSE, we will start by reviewing standard electroluminescence in polaritonic systems for $\eta\ll1$ \cite{DeLiberato08,DeLiberato09b}. 
In this regime, the Rabi Hamiltonian reduces to a Jaynes-Cummings model
\begin{equation}
\label{JC}
\ket{G}\simeq\ket{g,0},\quad\quad \ket{\pm}\simeq\tfrac{\ket{g,1}\pm\ket{e,0}}{\sqrt{2}},
\end{equation}
leading to $\hbar\omega_G=0$ and $\Gamma_{\text{out}}^{G\rightarrow n}=\Gamma_{\text{out}}\delta_{n,0}$,
$\Gamma_{\text{in}}^{n\rightarrow G}=\Gamma_{\text{in}}\delta_{n,0}$: as the ground state $\ket{G}$ contains no photons, when we extract the electron we cannot end in a state with a photonic component and vice versa.
A current can thus pass through the structure for any $\mu\geq0$, without any photonic emission, through the path
\begin{equation}
\label{GSE00}
\ket{s,0}\xrightarrow{\Gamma_{\text{in}}}\ket{G}\xrightarrow{\Gamma_{\text{out}}}\ket{s,0}.
\end{equation}
Only for $\mu\geq\hbar\omega_{G}+\hbar\omega_{\pm}$, the electrons can excite states with a photonic component, corresponding to $\Gamma_{\text{in}}^{0\rightarrow\pm}>0$, then decaying to the ground state before being extracted. This leads to the usual polaritonic electroluminescence at energy $\hbar\omega_{\pm}$, through the processes
\begin{equation}
\label{EL1}
\ket{s,0}\xrightarrow{\Gamma_{\text{in}}^{0\rightarrow \pm}}\ket{\pm}\xrightarrow[{\rotatebox[origin=c]{-90}{$\rightarrow$}}\,\,\,\hbar\omega_{\pm}]{\Gamma_{\text{cav}}^{\pm}}\ket{G}\xrightarrow{\Gamma_{\text{out}}}\ket{s,0},
\end{equation}
where we highlighted the transition (and its energy) expected to emit light.

The situation radically changes when instead $\eta\simeq 1$. In this case Eq.~\ref{JC} is not valid anymore, and the ground state has a non-negligible photonic component, $\bra{G}a^{\dagger}a\ket{G}\neq0$ \cite{Ciuti05}.
As $\ket{G}$ is now a linear superposition of states with one electron and $n$ photons, when the electron is extracted, it has a finite probability to project the system in the state $\ket{s,n}$, with no electrons and $n$ photons ($\Gamma_{\text{out}}^{G\rightarrow n}>0$), that will eventually escape out of the cavity.
This process is what we term GSE: emission of photons when a current passes through the ground state. The fundamental GSE process, emitting at the cavity frequency $\hbar\omega_C$, has the form
\begin{equation}
\label{GSE1}
\ket{s,0}\xrightarrow{\Gamma_{\text{in}}^{0\rightarrow G}}\ket{G}\xrightarrow{\Gamma^{G\rightarrow 1}_{\text{out}}}\ket{s,1}
\xrightarrow[{\rotatebox[origin=c]{-90}{$\rightarrow$}}\,\,\,\hbar\omega_{C}]{\Gamma_{\text{cav}}}\ket{s,0}.
\end{equation}
If the photonic lifetime is longer than the electron injection and extraction times, GSE photons can accumulate in the cavity, leading to the occupation of higher-lying states of the Rabi Hamiltonian and luminescence at the respective frequencies. Under the hypothesis of large enough cavity losses, we will limit ourselves to processes of first order in $\tfrac{\Gamma_{\text{in/out}}}{\Gamma_{\text{cav}}}$
\begin{eqnarray}
\label{GSE2}&&
\ket{s,0}\xrightarrow{\Gamma_{\text{in}}^{0\rightarrow G}}\ket{G}\xrightarrow{\Gamma^{G\rightarrow 1}_{\text{out}}}\ket{s,1}\xrightarrow{\Gamma^{1\rightarrow\pm}_{\text{in}}}\ket{\pm}
\xrightarrow[{\rotatebox[origin=c]{-90}{$\rightarrow$}}\,\,\,\hbar\omega_{\pm}]{\Gamma_{\text{cav}}^{\pm}}\ket{G}\nonumber\\&&
\xrightarrow{\Gamma^{G\rightarrow0}_{\text{out}}}\ket{s,0},
\end{eqnarray}
that clearly emit  at the polaritonic frequencies $\hbar\omega_{\pm}$.
We emphasize that the processes in Eqs.~(\ref{GSE1}, \ref{GSE2}) do not depend upon $\Gamma^{0\rightarrow \pm}_{\text{in}}$, and thus from \Eq{Inrate} they happen for values of the applied bias $\mu<\hbar\omega_G+\hbar\omega_{-}$, such that the standard electroluminescence in Eq.~(\ref{EL1}) is forbidden.

\begin{figure}
\centering
\begin{tikzpicture}

\node at (0,0.){\includegraphics[width=\linewidth]{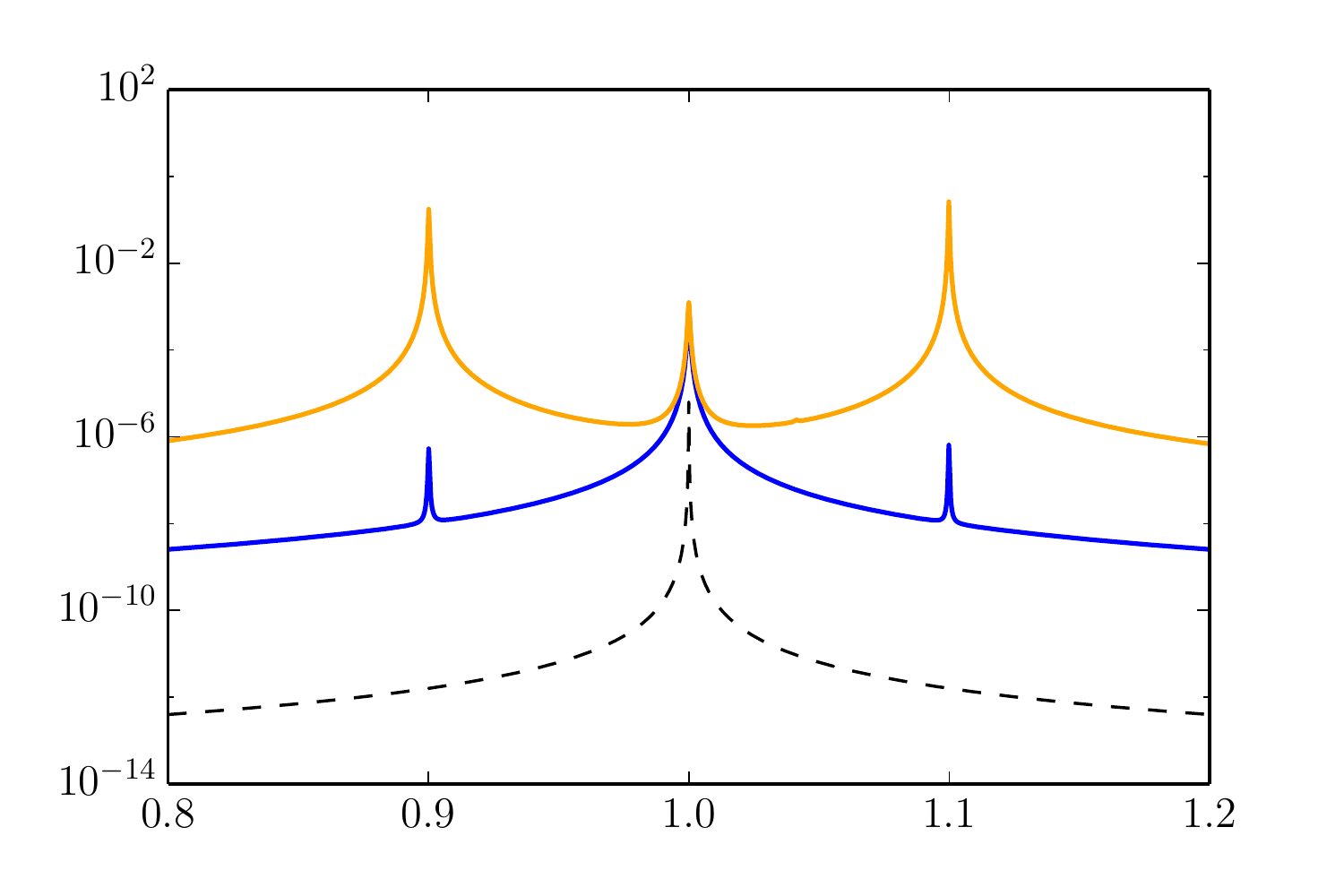}};
\node [font=\large,scale=1.1,style=black] at  (0.1,1.5) {C};
\node [font=\large,scale=1.1,style=black] at  (1.8,1.8) {$+$};
\node [font=\large,scale=1.1,style=black] at  (-1.6,1.8) {$-$};
\node [font=\large,scale=1,style=black] at  (0.1,-3) {$\omega$ (in units of $\omega_C$)};
\node [font=\large,scale=1,style=black,rotate=90] at  (-4.4,0.1) {$S(\omega)$};
  \end{tikzpicture}  
  \caption{\label{Fig2} Emission spectrum $S(\omega)$ for GSE alone (blue solid line, $\mu=\hbar\omega_G$) and GSE plus standard electroluminescence (yellow solid line, $\mu=\hbar\omega_+$). The peaks corresponding to the processes emitting at the bare cavity frequency $\hbar\omega_C$ and at the polaritonic frequencies $\hbar\omega_{\pm}$ are explicitly shown. The dashed line represents the black-body radiation spectrum for $k_{\text{B}} T\sim 0.1~ \hbar \omega_C$.
Parameters: $\eta=0.1$, $\Gamma=0.5 \times 10^{-6}\omega_C$, $\Gamma_{\text{cav}}=7\times10^{-4}\omega_C$.}
\end{figure}
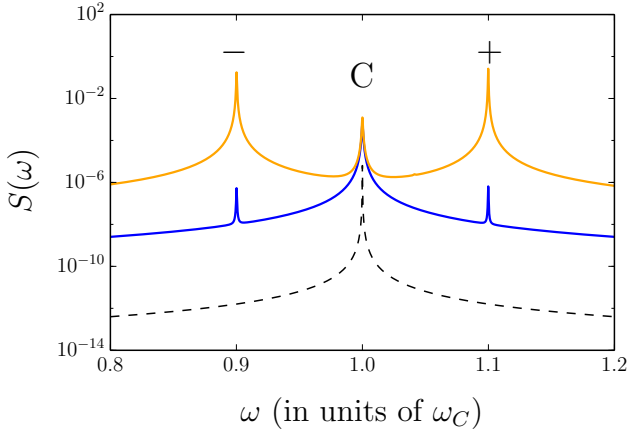

The GSE emission rate can be calculated writing a rate equation for the populations $P_j$ of the respective states $\ket{j}$. Limiting ourselves to the first satellite peaks, and to the regime $\mu<\hbar\omega_G+\hbar\omega_-$, in which no standard electroluminescence is possible, we have
\begin{equation}
\begin{array}{cll}
\dot{P}_{s,0}&=&-P_{s,0} \Gamma^{s,0}_{\text{in}}+P_{G} \Gamma_{\text{out}}^{G\rightarrow0}
+P_{\pm} \Gamma_{\text{out}}^{\pm\rightarrow0}
+P_{s,1}\Gamma_{\text{cav}}\\

\dot{P}_{s,1}&=&-P_{s,1} (\Gamma_{\text{cav}}+\Gamma^{s,1}_{\text{in}})+P_{G}\Gamma_{\text{out}}^{G\rightarrow1}+P_{\pm}\Gamma_{\text{out}}^{\pm\rightarrow1}\\

\dot{P}_G&=&-P_{G}\Gamma^G_{\text{out}}+P_{s,0} \Gamma_{\text{in}}^{0\rightarrow G}+P_{s,1} \Gamma_{\text{in}}^{1\rightarrow G}+P_{\pm}\Gamma_{\text{cav}}^{\pm}\\

\dot{P}_{+}&=&-P_{+}(\Gamma_{\text{cav}}^{+}+	
\Gamma^+_{\text{out}})+P_{s,1}\Gamma_{\text{in}}^{1\rightarrow+}\\
\dot{P}_{-}&=&-P_{-}(\Gamma_{\text{cav}}^{-}+
\Gamma^-_{\text{out}})+P_{s,1}\Gamma_{\text{in}}^{1\rightarrow-}\;,
\end{array}
\end{equation}
where repeated $\pm$ stands for sum over its repeated constituents. In the steady state, this system can be solved for the populations and, consequently, the emission rates for the fundamental GSE in \Eq{GSE1}, $f_C$, and for the first satellites in \Eq{GSE2},  $f_{\pm}$, can  be calculated as
\begin{equation}
f_C=P_{s,1}\Gamma_{\text{cav}},\quad f_{+}=P_{+}\Gamma_{\text{cav}}^{+},\quad f_{-}=P_{-}\Gamma_{\text{cav}}^{-}.
\end{equation}
Developing perturbatively the coupled eigenstates of the Rabi Hamiltonian $\ket{G}$ and $\ket{\pm}$ over the uncoupled states $\ket{g,n}$ and $\ket{e,n}$, and considering $\Gamma_{\text{in}}=\Gamma_{\text{out}}\equiv\Gamma$, we can obtain estimates for the radiation intensity of the main GSE peak in \Eq{GSE1} and for the satellites from \Eq{GSE2}. To the lowest nontrivial order in $\eta$ and to the first order in
$\frac{\Gamma}{\Gamma_{\text{cav}}}$ they read
\begin{eqnarray}
\label{eq:f}
f_C&=&\frac{\eta^2\Gamma}{8}\left(1-\frac{\Gamma}{\Gamma_{\text{cav}}}\right),\quad
f_{\pm}=\frac{\eta^2\Gamma}{16}\frac{\Gamma}{\Gamma_{\text{cav}}},
\end{eqnarray}
where, as expected, the emission at $\hbar\omega_{\pm}$, that relies upon a photon buildup in the cavity as from \Eq{GSE2}, is weighted down by a factor $\tfrac{\Gamma}{\Gamma_{\text{cav}}}\ll 1$.

For comparison we also calculated, to the same order, the standard electroluminescence intensity for the case $\mu\geq\hbar\omega_G+\hbar\omega_+$, obtaining
\begin{eqnarray}
\label{eq:f0}
f'_C=\frac{\Gamma}{6}\left(\frac{2\Gamma}{\Gamma_{\text{cav}}}+\eta^2\right), ~ f'_{\pm}=\frac{\Gamma}{6} \left(1\pm\frac{\eta}{2}\right)\left(1-\frac{2\Gamma}{\Gamma_{\text{cav}}}\right).
\end{eqnarray}
To the dominant order in both variables we obtain, as expected, a term not dependent on $\eta$ emitting at $\hbar\omega_{\pm}$, that is the standard electroluminescence described in \Eq{EL1}. Note that also in this case we recover higher-order processes allowed by the buildup of photonic population in the cavity (and thus of higher order in $\tfrac{\Gamma}{\Gamma_{\text{cav}}}$);  in this case the first one emitting at $\hbar\omega_C$
\begin{equation}
\label{EL2}
\ket{s,0}\xrightarrow{\Gamma_{\text{in}}^{0\rightarrow \pm}}\ket{\pm}\xrightarrow{\Gamma_{\text{out}}^{\pm\rightarrow1}}\ket{s,1}\xrightarrow[{\rotatebox[origin=c]{-90}{$\rightarrow$}}\,\,\,\hbar\omega_{C}]{\Gamma_{\text{cav}}}\ket{s,0}.
\end{equation}

\begin{figure}
\centering
\begin{tikzpicture}

\node at (-6,-6){\includegraphics[width=\linewidth]{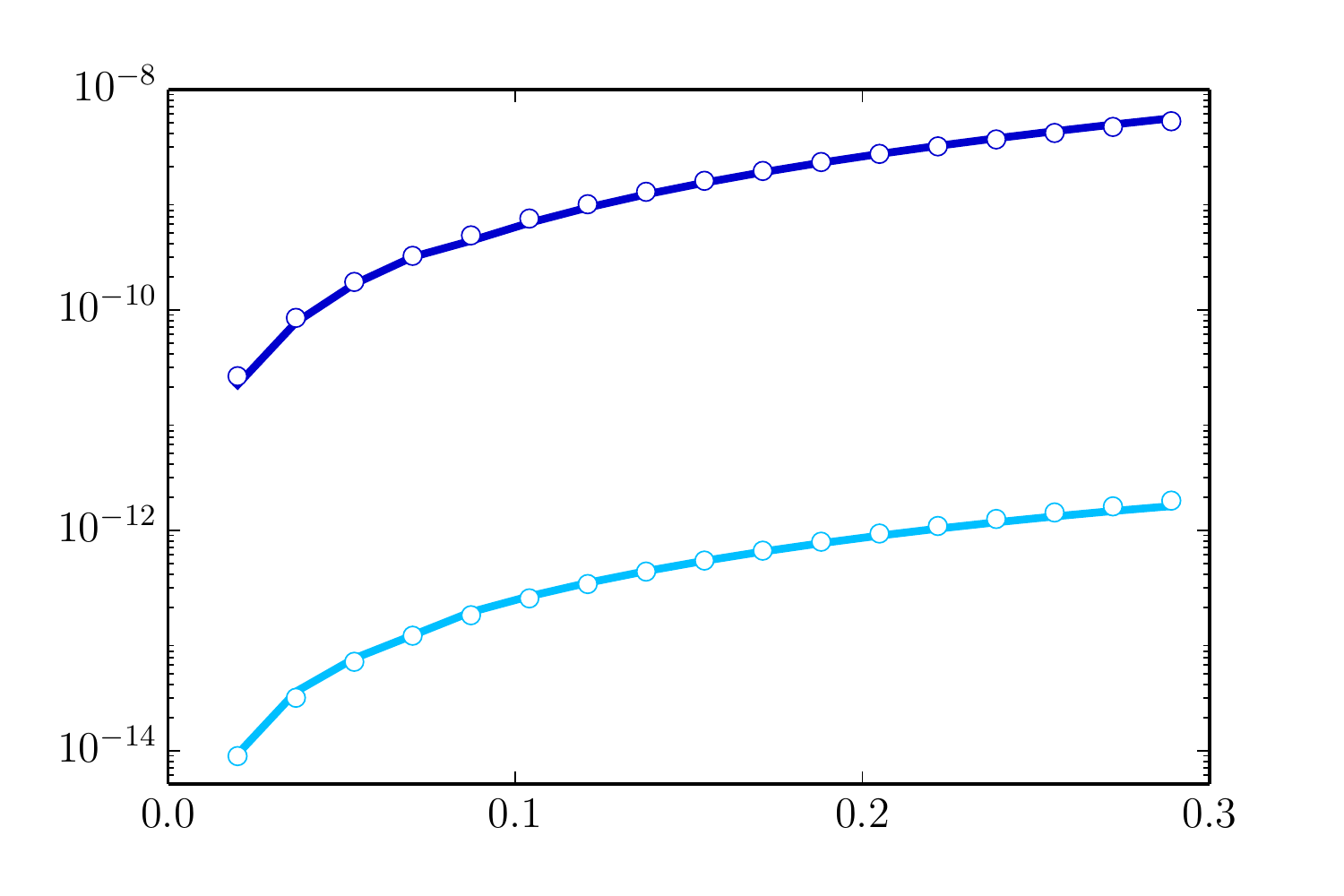}};
\node at (-6,-11.4){\includegraphics[width=\linewidth]{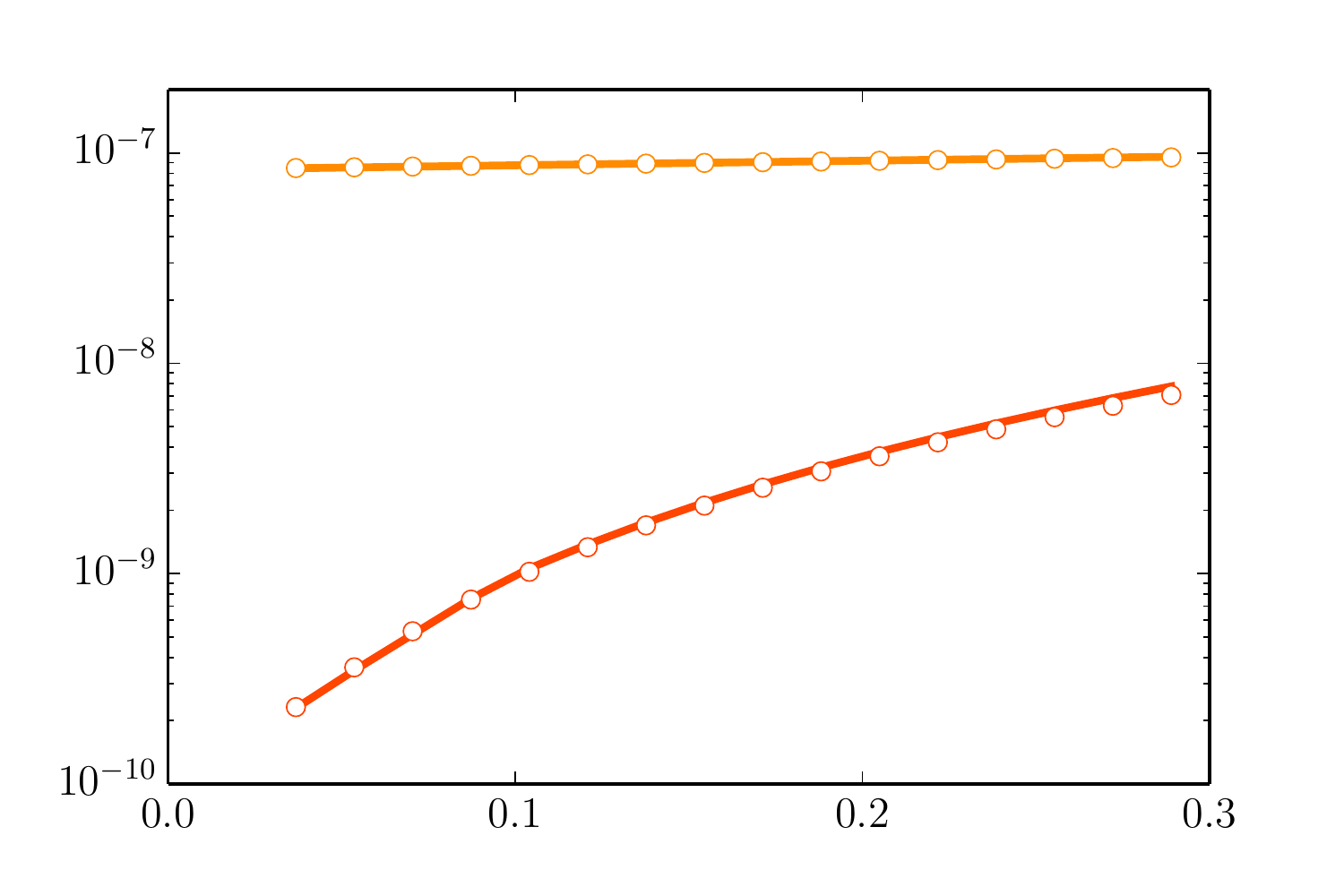}};
                 
\node at (-7.5,-11){\scalebox{0.52}{\begin{tikzpicture}[>=latex,thick,font=\normalsize]
\node at (0,0)[rectangle,draw=black!100,fill=black!100,thick,inner sep=0pt,minimum width=1.8cm,minimum height=.01cm] (a) {};
\node at (0,-1.5)[rectangle,draw=black!100,fill=black!100,thick,inner sep=0pt,minimum width=1.8cm,minimum height=.01cm] (b) {};
\node at (0,-4.)[rectangle,draw=black!100,fill=black!100,thick,inner sep=0pt,minimum width=1.8cm,minimum height=.01cm] (c) {};
\node at (2,-2.9)[rectangle,draw=black!100,fill=black!100,thick,inner sep=0pt,minimum width=1.8cm,minimum height=.01cm] (d) {};
\draw [-,RedOrange,shorten >=2pt,decoration={markings, mark=at position 0.999 with {\arrow[scale=1.5]{>}}; }, postaction={decorate}] ($($(c)$)-(1.,-0.28)$) -- ($($(a)$)-(1.,0)$); 
\draw [-,RedOrange2,shorten >=2pt,decoration={markings, mark=at position 0.999 with {\arrow[scale=1.5]{>}}; }, postaction={decorate}] ($($(a)$)+(1.4,-0.48)$) -- ($($(d)$)-(-.8,-0.2)$); 
\draw [-,RedOrange,shorten >=2pt,decoration={markings, mark=at position 0.999 with {\arrow[scale=1.5]{>}}; }, postaction={decorate}] ($($(b)$)+(0.8,-0.33)$) -- ($($(c)$)+(0.8,0.05)$); 
\node [scale=0.7, single  arrow, draw,rotate=-135,draw=black!60,fill=RedOrange2!80] at ($($(d)$)-(.6,0.6)$) {~~~~~~~~~~}; 
\node [scale=0.7, single  arrow, draw,rotate=-90,draw=black!60,fill=RedOrange1!80] at ($($(a)$)-(-0.8,0.7)$) {~~~~~~~~~~}; 
\node [font=\LARGE,scale=1,style=RedOrange1] at (1.2,-0.7) {+};
\node [font=\LARGE,scale=1,style=RedOrange2] at ($($(d)$)-(0.2,0.7)$) {C};
\node [font=\large,scale=1.1,style=black] at  ($($(c)$)-(0.3,-0.35)$) {$\ket{s,0}$};
\node [font=\large,scale=1.1,style=black] at  ($($(b)$)-(0.3,-0.35)$) {$\ket{G}$};
\node [font=\large,scale=1.1,style=black] at  ($($(a)$)-(0.3,0.35)$) {$\ket{+}$};
\node [font=\large,scale=1.1,style=black] at  ($($(d)$)-(0.3,-0.35)$) {$\ket{s,1}$};
\end{tikzpicture}}};

\node at (-3.7,-5.26){\scalebox{0.52}{\begin{tikzpicture}[>=latex,thick,font=\normalsize]
\node at (0,0)[rectangle,draw=black!100,fill=black!100,thick,inner sep=0pt,minimum width=1.8cm,minimum height=.01cm] (a) {};
\node at (0,-1.5)[rectangle,draw=black!100,fill=black!100,thick,inner sep=0pt,minimum width=1.8cm,minimum height=.01cm] (b) {};
\node at (0,-4.)[rectangle,draw=black!100,fill=black!100,thick,inner sep=0pt,minimum width=1.8cm,minimum height=.01cm] (c) {};
\node at (2,-2.9)[rectangle,draw=black!100,fill=black!100,thick,inner sep=0pt,minimum width=1.8cm,minimum height=.01cm] (d) {};
\draw [-,Blue1,shorten >=2pt,decoration={markings, mark=at position 0.999 with {\arrow[scale=1.5]{>}}; }, postaction={decorate}] ($($(c)$)-(1.,-0.28)$) -- ($($(b)$)-(1.,0)$); 
\draw [-,Blue2,shorten >=2pt,decoration={markings, mark=at position 0.001 with {\arrow[scale=1.5]{<}}; }, postaction={decorate}] ($($(a)$)+(1.4,-0.48)$) -- ($($(d)$)-(-.8,-0.2)$); 
\draw [-,Blue1,shorten >=2pt,decoration={markings, mark=at position 0.999 with {\arrow[scale=1.5]{>}}; }, postaction={decorate}] ($($(b)$)+(1,-0.33)$) -- ($($(b)$)+(1,-1.3)$); 
\draw [-,Blue2,shorten >=2pt,decoration={markings, mark=at position 0.001 with {\arrow[scale=1.5]{<}}; }, postaction={decorate}] ($($(c)$)+(.8,0.33)$) -- ($($(b)$)+(.8,-0.05)$); 
\node [scale=0.7, single  arrow, draw,rotate=-135,draw=black!60,fill=Blue1!80] at ($($(d)$)-(.6,0.6)$) {~~~~~~~~~~}; 
\node [scale=0.7, single  arrow, draw,rotate=-90,draw=black!60,fill=Blue2!80] at ($($(a)$)-(-0.8,0.7)$) {~~~~~~~~~~}; 
\node [font=\LARGE,scale=1,style=Blue2] at (1.2,-0.7) {+};
\node [font=\LARGE,scale=1,style=Blue1] at ($($(d)$)-(0.2,0.7)$) {C};
\node [font=\large,scale=1.1,style=black] at  ($($(c)$)-(0.3,-0.35)$) {$\ket{s,0}$};
\node [font=\large,scale=1.1,style=black] at  ($($(b)$)-(0.3,-0.35)$) {$\ket{G}$};
\node [font=\large,scale=1.1,style=black] at  ($($(a)$)-(0.3,0.35)$) {$\ket{+}$};
\node [font=\large,scale=1.1,style=black] at  ($($(d)$)-(0.3,-0.35)$) {$\ket{s,1}$};
\end{tikzpicture}}};

                    \node [font=\large,scale=1.1,style=black] at  (-6,-14.3) {$\eta$};
                    \node [font=\large,scale=1.1,style=RedOrange2] at  (-3.35,-10.8) {C};
                    \node [font=\large,scale=1.1,style=RedOrange1] at  (-3.35,-9.32) {+};
                    \node [font=\large,scale=1.1,style=Blue2] at  (-8,-7) {+};
                    \node [font=\large,scale=1.1,style=Blue1] at  (-8,-4.5) {C};

                      \node [font=\large,scale=.8,style=black,rotate=90] at  (-10.3,-6.) {$f_C~,~ f_+$ (in units of $\omega_C$)};
                     \node [font=\large,scale=.8,style=black,rotate=90] at  (-10.3,-11.2) {$f^\prime_C~,~ f^\prime_+$ (in units of $\omega_C$)};
                     
                     \node [font=\large,scale=.8,style=black] at  (-8.9,-4) {(a)};
                       \node [font=\large,scale=.8,style=black] at  (-8.9,-9.43) {(b)};
                     
  \end{tikzpicture}  

\caption{\label{Fig3} (a) Integrated  emission spectrum $S(\omega)$ (as in Fig.~\ref{Fig2}) for the central peak and one of the satellites channels for $\mu=\hbar\omega_G$, obtained by integrating the numerical spectrum (solid lines) and from the analytical estimates in \Eq{eq:f} (hollow dots). In the inset we illustrate the GSE processes from Eqs.~(\ref{GSE1},\ref{GSE2}).
 (b) Integrated emissions for $\mu=\hbar\omega_G+\hbar\omega_+$, obtained by integrating the numerical spectrum (solid lines), and from the analytical estimates in Eq.~(\ref{eq:f0}) (hollow dots). Inset:  electroluminescence  from  Eqs.~(\ref{EL1},\ref{EL2}).}
\end{figure}
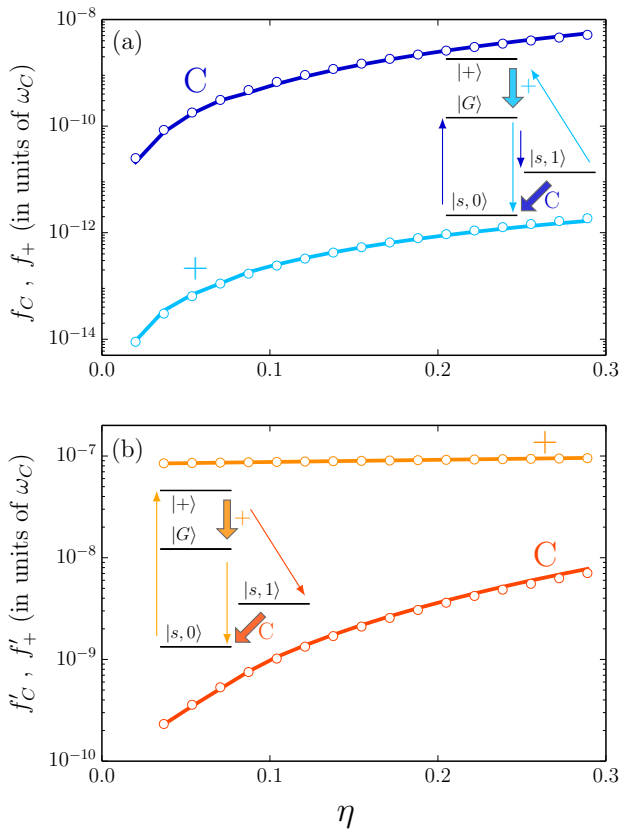

In order to test such an intuitive understanding of GSE, we studied the system numerically solving \cite{JohassonQutip1,JohassonQutip2} the master equation
\begin{equation}
\dot{\rho}=-i[H,\rho]+\mathcal{L}_{\text{in}}(\rho)+\mathcal{L}_{\text{out}}(\rho)+\mathcal{L}_{\text{cav}}(\rho),
\end{equation}
where $\rho$ is the density operator of the system with a converged cutoff over the maximum number of photons and the Lindblad operators $\mathcal{L}_{\text{in}/\text{out}/\text{cav}}$ are projected over the dressed basis of the system \cite{DeLiberato09,Beaudoin11}.
Exploiting the input-output formalism compatible with the dressed state analysis developed in Ref. \cite{Ridolfo12}, we can write the extra-cavity emission spectrum as
\begin{equation}
\label{S}
S(\omega)=\frac{\Gamma_{\text{cav}}}{2\pi}\int_{-\infty}^{\infty} d\tau~ e^{-i\omega\tau}\langle X^+(\tau) X^-(0)\rangle,
\end{equation}
where $X=(a+a^\dagger)$, $X^-=\sum_{j>i}\bra{i}X\ket{j}\ket{i}\bra{j}$, and $X^+$ is its hermitian conjugate.
In Fig.~\ref{Fig2} we plot the emitted spectrum from \Eq{S} for $\mu=\hbar\omega_G$ (blue solid line) in which we can clearly see the central GSE peak at $\hbar\omega_C$ and the satellites around $\hbar\omega_{\pm}\simeq\hbar\omega_C\pm\Omega$. Figure \ref{Fig2}  also plots (yellow solid line) the spectrum for $\mu=\hbar\omega_G+\hbar\omega_+$. In this case, while the central peak remains substantially unchanged, standard electroluminescence becomes dominant over the satellite frequencies. There we plot, for comparison, the black body emission from the cavity for $k_{\text{B}}T=0.1\hbar\omega_C$.
In Fig.~\ref{Fig3} we plot (solid lines) the integrated emission of the central peak and of the two satellites respectively for $\mu=\hbar\omega_G$ (a) and $\mu=\hbar\omega_G+\hbar\omega_+$ (b), as a function of the perturbative parameter $\eta$.  The hollow dots in the same Figures are obtained by the analytical estimates from \Eq{eq:f} and \Eq{eq:f0}, respectively. Schematics of the involved processes are shown as insets. We clearly obtain a very good quantitative agreement between our theory and the numerical simulations, proving the correctness of our theoretical analysis.

Various developing technological platforms seem possible candidates to experimentally observe GSE and some of them, like hybrid QED \cite{Liu14,Liu15,Deng14,Stockklauser15,Lambert13,Lambert15,Xiang13,Delbecq2011,Delbecq2013,Frey2012,Frey2012b,Petersson2012,Toida2013,Wallraff2013,Viennot2014,Savage1992,McKeever2003} and circuit QED \cite{Baust15,Niemczyk10,Astafiev07,Ashhab09,Rodrigues07,You07,Hauss08,Devoret07,You05,You11}, in which electronic transitions are coupled to microwave superconducting resonators, seem close to achieve all the needed requirements. 
In particular, a double quantum dot coupled to a superconducting resonator seems to us the system better described by the simple model studied here, where $\ket{s}$ is the empty state, $\ket{g}$ and $\ket{e}$ are the hybridised states of the two dots, and the doubly-occupied state is Coulomb blocked. Remarkably, electroluminescence has already been observed in those systems \cite{Liu14,Stockklauser15}. Given that the ratio $\tfrac{\Gamma}{\Gamma_{\text{cav}}}$ can be engineered in those systems, from  \Eq{eq:f} we obtain a rough estimate of the photon population, due to GSE, of $P\simeq 0.1\eta^2$. Considering a conservative lower bound to the detectable photon population in microwave resonators of $P_\text{min}=10^{-3}$ \cite{Stockklauser15}, $\eta>0.1$ should thus suffice to observe GSE. This normalised coupling has been achieved in circuit QED \cite{Niemczyk10,Baust15}. In hybrid systems, experiments are moving toward higher couplings and the appropriate regime may become accessible in the near future.

In conclusion, we have highlighted a novel cavity QED phenomenon, leading to the purely quantum emission of photons out of the ground state when an electric current passes through it. Such a novel form of quantum vacuum emission could be observed in various near-future cavity QED setups, leading to a further step forward in the understanding and engineering of the quantum vacuum in the ultrastrong light-matter coupling regime.\\

\textbf{Acknowledgments.}- This work is partially supported by the RIKEN iTHES Project, the MURI Center for Dynamic Magneto-Optics
via the AFOSR award number FA9550-14-1-0040, the IMPACT program of JST, and a Grant-in-Aid for Scientific Research (A). MC is supported by the Canon Foundation in Europe and the RIKEN iTHES program. NL is partially supported by the FY2015 Incentive Research Project. SDL  is Royal Society Research Fellow. SDL acknowledges support from the Engineering and Physical Sciences Research Council (EPSRC), research grant EP/L020335/1.

\bibstyle{plain}

\end{document}